\documentclass[conference]{IEEEtran}
\IEEEoverridecommandlockouts
\usepackage{amsmath,amsfonts}
\usepackage{mathtools}
\usepackage{amsthm}
\usepackage{amssymb}
 \usepackage{multirow}
\usepackage{array}
\usepackage{textcomp}
\usepackage{stfloats}
\usepackage{url}
\usepackage{microtype}
\usepackage{verbatim}
\usepackage{graphicx}
\usepackage{booktabs}
\usepackage{cite}
\usepackage[T1]{fontenc}
\usepackage[colorlinks=true, linkcolor=blue, citecolor=blue, urlcolor=blue]{hyperref}
\usepackage{enumitem}

\usepackage[linesnumbered,lined,algoruled,commentsnumbered]{algorithm2e}

\usepackage{caption}
\usepackage{subcaption}
\usepackage{stfloats}
\usepackage{lipsum} 
\usepackage[font=small,skip=0pt]{caption}
\usepackage[export]{adjustbox}
\usepackage[inkscapelatex=false]{svg}

\SetCommentSty{mycommfont}
\theoremstyle{definition}

\usepackage{units}
\newcommand{\nth}[1]{{#1}^{\text{th}}}

\newcommand{\abs}[1]{\left|{#1}\right|}

\newcommand{\RD}[0]{r_{\mathrm{\scalebox{0.5} {RD} }}}

\newcommand{\rf}[0]{r_{\mathrm{\scalebox{0.5} {0}}}}

\newcommand{\GUCA}[0]{\mathcal{G}_\mathrm{\scalebox{0.5} {uca}}}
\newcommand{\GULA}[0]{\mathcal{G}_\mathrm{\scalebox{0.5} {ula}}}
\newcommand{\GURA}[0]{\mathcal{G}_\mathrm{\scalebox{0.5} {ura}}}
\newcommand{\GUCCA}[0]{\mathcal{G}_\mathrm{\scalebox{0.5} {ucca}}}
\newcommand{\RDUCA}[0]{{r}_\mathrm{ \scalebox{0.5} {RD}}^{\scalebox{0.5} {uca}}}
\newcommand{\DUCA}[0]{D_{\scalebox{0.5} {uca}}}
\newcommand{\RDUCCA}[0]{{r}_\mathrm{ \scalebox{0.5} {RD}}^{\scalebox{0.5} {ucca}}}
\newcommand{\DUCCA}[0]{D_{\scalebox{0.5} {ucca}}}

\usepackage{acronym}
\input{acronym}

\begin{document}
\title{
Array Geometry-Centric Axial Sidelobe Interference Analysis for Near-Field Multi-User MIMO\\ 

\author{\IEEEauthorblockN{Ahmed Hussain\IEEEauthorrefmark{1}, Asmaa Abdallah\IEEEauthorrefmark{1}, Abdulkadir Celik\IEEEauthorrefmark{2}, Ahmed M. Eltawil\IEEEauthorrefmark{1}}
\IEEEauthorblockA{\IEEEauthorrefmark{1}King Abdullah University of Science and Technology (KAUST), Thuwal, Saudi Arabia}
\IEEEauthorblockA{\IEEEauthorrefmark{2}School of Electronics and Computer Science, University of Southampton, SO17 1BJ UK}
}
}

\maketitle

\begin{abstract}
With the deployment of large antenna arrays at high-frequency bands, future wireless communication systems are likely to operate in the radiative near-field (NF). Unlike far-field beam steering, NF beams can be focused on a spatial region with finite depth, enabling user multiplexing in both range and angle. In NF multiuser multiple-input multiple-output (MU-MIMO) systems, achievable rates are limited by interference arising from sidelobes in both the axial (range) and lateral (angle) dimensions. This work investigates how axial sidelobes (ASLs) vary with array geometry. Closed-form array gain expressions are derived to characterize ASLs for uniform planar arrays. Analytical results show that the uniform square array (USA) yields the lowest ASLs, followed by the uniform concentric circular array (UCCA), uniform linear array (ULA), and uniform circular array (UCA). Specifically, the USA achieves a peak sidelobe level (PSLL) of $-17.6$ dB versus $-7.9$ dB for the UCA. Numerical simulations confirm that the USA provides superior sidelobe suppression and highest sumrate performance.
\end{abstract}
\begin{IEEEkeywords} Near field, UM-MIMO, array geometry, axial sidelobes, sidelobe level, ULA, URA, UCA, UCCA.
\end{IEEEkeywords}

\section{Introduction} \label{Sec_intro}
\IEEEPARstart{W}{ith} the growing adoption of massive \ac{MIMO} in \ac{5G} communications, future networks are anticipated to embrace \ac{UM}-\ac{MIMO} arrays operating at higher frequency bands. A higher carrier frequency implies a shorter wavelength and therefore a smaller antenna size. As the array aperture increases and the wavelength decreases, the radiative \ac{NF} region expands significantly, since it is proportional to the squared aperture length and inversely proportional to the wavelength. Consequently, future communication systems are also likely to operate within the \ac{NF} region.

Wave propagation in the radiative \ac{NF} is characterized by spherical wavefronts, in contrast to the planar wavefront assumption of the \ac{FF}. Unlike classical \ac{FF} beamforming, which steers beams only toward angular directions, \ac{NF} beamforming can focus energy at specific locations in both angle and range, similar to a spotlight—a phenomenon known as \ac{NF} beamfocusing \cite{11095387}. Consequently, \ac{NF} \acp{UE} can be multiplexed across both angular and range domains, thereby improving spatial multiplexing gains.

The \ac{MU}-\ac{MIMO} communication is limited by interference arising from both the mainlobe and sidelobes. The resolution of the mainlobe in the range dimension is characterized by the beamdepth. The mutual interference between \acp{UE} can be mitigated by reducing the beamdepth, which depends on the array geometry \cite{10934779}. Under the constraint of a fixed number of antenna elements, elongated \acp{URA}, such as a \ac{ULA}, provide the narrowest beamdepth, followed by \acp{UCCA} and \acp{USA} \cite{10443535}. In a \ac{ULA}, the beamdepth is narrowest at boresight and broadens toward endfire directions \cite{10570663,10988573}, thereby confining \ac{NF} benefits primarily to \acp{UE} near boresight. To enhance angular coverage, \acp{UCA} have been proposed due to their rotational symmetry \cite{10243590}, but they cannot form finite-depth beams at boresight. This limitation is addressed by \acp{UCCA}, which enable finite-depth \ac{NF} beams at boresight while slightly reducing \ac{NF} coverage in coplanar directions \cite{10517927}. Consistent with these observations, the impact of array geometry on beamdepth is pronounced under a fixed element count constraint, but becomes less significant when the aperture length is fixed \cite{ahmed2025NF}.

\begin{figure}[t]
\centering
\includegraphics[width=1\columnwidth]{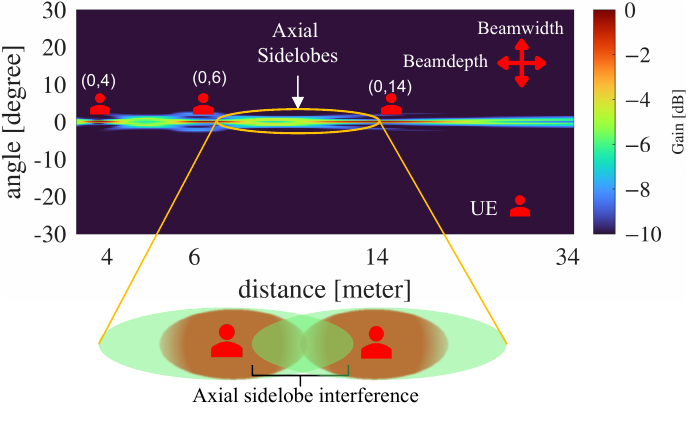}
\setlength{\belowcaptionskip}{-20pt}
\caption{Interference due to axial sidelobes in \ac{MU}-\ac{MIMO}.}
\label{fig:MU-MIMO interfernce}
\end{figure}
When \acp{UE} are separated by distances smaller than the mainlobe resolution (beamwidth or beamdepth), interference arises from both the mainlobe and sidelobes; for larger separations, interference is predominantly due to sidelobes as indicated by the non-overlapping beams in Fig. \ref{fig:MU-MIMO interfernce}. Unlike the \ac{FF}, where interference originates exclusively from \acp{LSL}, \ac{NF} beams exhibit both \acp{ASL} and \acp{LSL}, making them susceptible to interference in both angular and range domains. The overall pattern in the angle domain is determined by the product of the element factor and the array factor; reducing element sidelobes directly lowers overall array sidelobes \cite{balanis2016antenna}. In contrast, the element factor has no impact on the beam pattern in the range domain. This distinction underscores the importance of analyzing \acp{ASL} across the range domain and developing methods to suppress them.

Prior studies have characterized how beamdepth varies with array geometry \cite{10934779,10988573, 10570663,10443535,10243590,10517927,ahmed2025NF}. However, the behavior of \acp{ASL} across different geometries remains unexplored. To address this gap, we characterize the \ac{SLL} across various array geometries and \textit{identify the geometry that yields the lowest \acp{ASL}}. Specifically, we formulate the array gain function in the range domain and quantify \ac{SLL} with respect to \ac{PSL} and \ac{ISL}. We compare \ac{PSL} and \ac{ISL} for \ac{ULA}, \ac{URA}, \ac{UCA}, and \ac{UCCA}. Both theoretical analysis and numerical simulations show that \ac{USA} achieves the lowest \acp{ASL}, thereby providing the highest multiuser sumrate.

\section{System Model} \label{Sec_II}
We consider a \ac{MU}-\ac{MIMO} communication system, where a \ac{BS} equipped with $N$ isotropic antennas simultaneously serves $K$ single-antenna \acp{UE}. The received signal at the $\nth{k}$ \ac{UE} is expressed as
\begin{equation} 
y_{k}=\sqrt{\gamma}\,\mathbf{w}_{k} \mathbf{h}_{k} s_{k}
+ \sqrt{\gamma}\sum_{j=1, j \neq k}^{K}\mathbf{w}_{j}\mathbf{h}_{k} s_{j} 
+ z_{k}, 
\label{eqn_IIA_1}
\end{equation}
where $\gamma = P/N$ denotes the average transmit \ac{SNR}, $P$ is the total transmit power, $s_{k}$ is the unit-power transmit symbol, $\mathbf{w}_{k} \in \mathbb{C}^{1 \times N}$ is the precoding vector, and $z_{k}\!\sim\!\mathcal{CN}(0,1)$ represents complex Gaussian noise. The \ac{NF} channel vector $\mathbf{h}_{k} \in \mathbb{C}^{N \times 1}$ in \eqref{eqn_IIA_1} is modeled as
\begin{equation} 
\mathbf{h}_{k}=\beta_{k}\, \mathbf{b}(\varphi_{k},\theta_{k},r_{k} ), 
\label{eqn_IIA_2}
\end{equation}
where $\beta_{k}$ is the complex path gain and $\mathbf{b}(\cdot)$ is the \ac{NF} array response vector focused at azimuth angle $\varphi_{k}$, elevation angle $\theta_{k}$, and range $r_{k}$. In \ac{mmWave} systems, the \ac{LoS} path typically exceeds other multipath components by about $\unit[4\!-\!5]{dB}$. Hence, we focus on the \ac{LoS} path and set $\beta_k=1$, for simplicity. The normalized \ac{NF} array response vector in \eqref{eqn_IIA_2} is given by
\begin{equation} 
\mathbf{b}(\varphi,\theta,r) = \tfrac{1}{\sqrt{N}}\!\left[e^{-j\nu(r^{(0)} -r)}, \dots, e^{-j\nu(r^{(N-1)} -r)}\right]^\mathsf{T},
\label{eqn-A1}
\end{equation}
where $\nu = \tfrac{2\pi}{\lambda}$ is the wavenumber, $\lambda$ is the wavelength, and $r^{(n)}$ denotes the distance between the \ac{UE} and the $\nth{n}$ antenna.
\subsection{Achievable Rate}
The achievable rate for the $\nth{k}$ \ac{UE} is expressed as
\begin{equation}
\mathcal{R}_{k}=\log _{2}\!\left(1+\operatorname{SINR}_{k}\right),
\label{eqn_IIB_1}
\end{equation}
where, \ac{SINR} is
\begin{equation}
\operatorname{SINR}_{k}=\frac{\gamma|\mathbf{w}_{k}\mathbf{h}_{k}|^{2}}{1+\gamma \sum_{j=1, j \neq k}^{K}|\mathbf{w}_{j}\mathbf{h}_{k}|^{2}}. 
\label{eqn_IIB_2}
\end{equation}
Based on \eqref{eqn_IIB_1}, the achievable sumrate is given by $\mathcal{R}_{\mathrm{sum}}=\sum_{k=1}^{K} \mathcal{R}_{k}$. Assuming maximum-ratio transmission, the precoding vector is defined as $\mathbf{w}_{k}=\mathbf{h}_{k}^{\mathsf{H}}/\sqrt{N}$, where $(\cdot)^{\mathsf{H}}$ denotes the Hermitian operator. The maximum beamforming gain $\abs{\mathbf{w}_{k}\mathbf{h}_{k}}=N$, is achieved with the unnormalized \ac{NF} array response vector in \eqref{eqn-A1}. Accordingly, \eqref{eqn_IIB_2} is rewritten as
\begin{equation}
\mathcal{R}_{k}=\log _{2}\!\left(1+\frac{\gamma N}{1+\gamma N \sum_{j=1, j \neq k}^{K} \mathcal{I}_{a, kj}^{2}}\right), 
\label{eqn_IIB_3}
\end{equation}
where $a \in \{\mathrm{ula}, \mathrm{ura}, \mathrm{uca}, \mathrm{ucca}\}$ denotes the array configuration, and $\mathcal{I}_{a,kj}$ represents the interference term, expressed as the normalized cross-correlation between the desired and interfering channels 
\begin{equation}
\mathcal{I}_{a, kj}=\tfrac{1}{N}\big|\mathbf{h}_{k}^{\mathsf{H}}(\varphi_{k},\theta_{k},r_{k}) \mathbf{h}_{j}(\varphi_{j},\theta_{j},r_{j})\big|,
\label{eqn_B4}
\end{equation}
which evidently determines the achievable sumrate. Specifically, decreasing $\sum_{j=1,\, j \neq k}^{K} \mathcal{I}^2_{a, kj}$ suppresses inter-user interference and improves rate performance, whereas larger values of this term degrade the achievable rate. Moreover, \eqref{eqn_B4} can be analyzed in both angular and range domains. The angular-domain characteristics remain unchanged irrespective of whether the array operates in the \ac{NF} or \ac{FF} and have been extensively studied in prior work. In contrast, this study characterizes interference in the range domain.
\begin{figure}[t]
\centering
\includegraphics[width=1\columnwidth]{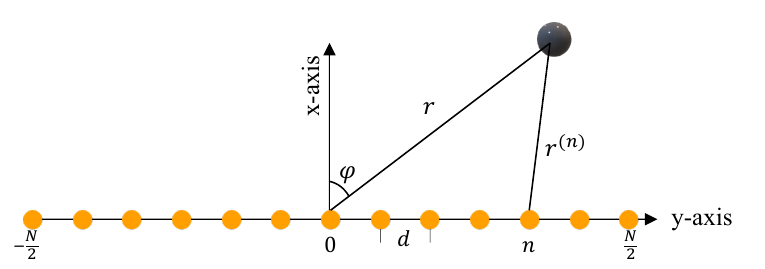}
\setlength{\belowcaptionskip}{-10pt}
\caption{A \ac{ULA} aligned along the $y$-axis and a \ac{NF} \ac{UE} at $(\varphi, r)$.}
\label{ula_system_model}
\end{figure}
\subsection{Axial Sidelobes}
In the downlink \ac{MU}-\ac{MIMO} system, the \ac{BS} simultaneously forms multiple \ac{NF} beams toward different \acp{UE}. Each beam comprises a mainlobe that delivers the desired signal and \acp{ASL} that cause inter-user interference. To quantify this sidelobe-induced interference, we formulate the axial beam pattern of the \ac{BS} focused on a single \ac{UE}. Assuming the \ac{UE} is located at a distance $\rf$ from the \ac{BS}, the axial beam pattern is given by
\begin{equation}
\mathcal{G}_a(\varphi,\theta,r) 
= \left|\mathbf{b}^\mathsf{H}(\varphi,\theta,\rf)\,\mathbf{b}(\varphi,\theta,r)\right|^2, 
\quad r \in [2D, \RD],
\label{eqn-A7r}
\end{equation}
where $D$ denotes the aperture length of the antenna array, and $\RD=\tfrac{2D^2}{\lambda}$ is the Rayleigh distance. The aperture length $D$ is defined as the largest dimension of the array. The minimum distance is set to $2D$ to ensure negligible amplitude variations across the array and employ a \ac{USW} model. The \acp{SLL} are evaluated using the following metrics:
\begin{enumerate}[label=\alph*)]
    \item \textbf{Peak Sidelobe Level}:  
    The \ac{PSL} quantifies the strongest sidelobe relative to the mainlobe and indicates the interference susceptibility from closely spaced users.  
    \begin{equation}
    \mathrm{PSLL} = 10 \log_{10} \!\left( \frac{\max_{r \in \mathcal{S}} \mathcal{G}_a(\varphi,\theta,r)}{\mathcal{G}_a(\varphi,\theta,\rf)} \right),
    \label{eqn-A8}
    \end{equation}
where $\mathcal{S}$ denotes the set of points in the sidelobe region.
    \item \textbf{Integrated Sidelobe Level}:  
    The \ac{ISL} measures the total sidelobe power relative to the mainlobe, reflecting the average interference experienced by all users.  
    \begin{equation}
    \mathrm{ISLL} = 10 \log_{10} \!\left( \frac{ \int_{\mathcal{S}} |\mathcal{G}_a(\varphi,\theta,r)|^2 \, dr }{ \int_{\mathcal{M}} |\mathcal{G}_a(\varphi,\theta,r)|^2 \, dr } \right),
    \label{eqn-A9}
    \end{equation}
where $\mathcal{M}$ denotes the set of points in the mainlobe region.
\end{enumerate}
\section{Axial Beam Pattern}\label{Sec-III}
In this section, the axial beam pattern is formulated, and the \ac{PSL} is derived for each array configuration. The analysis begins with the \ac{ULA}, as it directly extends to the \ac{URA}.
\subsection{Uniform Linear Array}
We consider a \ac{ULA} with $N$ antenna elements as shown in Fig. \ref{ula_system_model}. The \ac{UE} is located at a distance $r$ from the array center. The distance between \ac{UE} and the $\nth{n}$ antenna of the \ac{ULA}, obtained via the law of cosines, is
\begin{equation}
r^{(n)} = \sqrt{r^2 + n^2 d^2 - 2 r n d \sin(\varphi)},
\label{eqn_r_uca}
\end{equation}
where $\varphi$ denotes the azimuth angle. By substituting $r^{(n)}$ from \eqref{eqn_r_uca} into \eqref{eqn-A1}, the $\nth{n}$ component of the normalized \ac{NF} array response vector for the \ac{ULA} is given by
\begin{equation}
\mathbf{b}_n(\varphi,r) = \tfrac{1}{\sqrt{N}}
e^{-j\nu \left(\sqrt{r^2 + n^2 d^2 - 2 r n d \sin(\varphi)} - r \right)}.
\label{eqn-A3}
\end{equation}
Using a second-order Taylor expansion $\sqrt{1+t} \approx 1 + \tfrac{t}{2} - \tfrac{1}{8}t^2$, the distance difference is approximated as $r^{(n)} - r \approx n d \sin(\varphi) - \frac{n^2 d^2}{2r}\cos^2(\varphi)$ \cite{10517927}. Thus, the \ac{NF} array response vector in \eqref{eqn-A3} simplifies to
\begin{equation}
\mathbf{b}_n(\varphi,r) \approx \tfrac{1}{\sqrt{N}} 
e^{-j\nu \left(n d \sin(\varphi) - \tfrac{n^2 d^2}{2r}\cos^2(\varphi)\right)}.
\label{eqn-A4}
\end{equation}
Substituting \eqref{eqn-A4} into the general definition of the axial beam pattern in \eqref{eqn-A7r}, the array gain $\GULA$ is obtained as
\begin{align}
\GULA(\varphi,r) &= \left| \mathbf{b}^\mathsf{H}(\varphi,\rf)\,\mathbf{b}(\varphi,r) \right|^2, \label{eqn-B6-15}\\
&= \left|\frac{1}{N} \sum_{n=-N/2}^{N/2} e^{-j\nu n^2 d^2 \cos^2(\varphi) r_\mathrm{eff}} \right|^2, \label{eqn-B6-16}\\
\GULA(\gamma)&\approx \frac{C^2(\gamma) + S^2(\gamma)}{\gamma^2}, \label{eqn-B6-17}
\end{align}
where $r_\mathrm{eff} = \big| \tfrac{r-\rf}{2r\rf} \big|$. Introducing the variable transformation $x = \sqrt{\tfrac{n^2 d^2 \cos^2(\varphi)}{\lambda}\,\bigg| \tfrac{r-\rf}{r\rf} \bigg|}$, yields a Fresnel-integral representation of the array gain, with $\gamma = \sqrt{\tfrac{N^2 d^2 \cos^2(\varphi)}{\lambda} r_\mathrm{eff}}$, and $C(\gamma) = \int_0^\gamma \cos\!\left(\tfrac{\pi}{2}x^2\right)\,dx$, $S(\gamma) = \int_0^\gamma \sin\!\left(\tfrac{\pi}{2}x^2\right)\,dx$, are the Fresnel cosine and sine integrals, respectively. Moreover, $\gamma$ is a dimensionless parameter that captures the combined effects of array aperture, wavelength, angle, and the distance mismatch between the focusing point and the actual user position.

\begin{figure}[t]
\centering
\includegraphics[width=1\columnwidth]{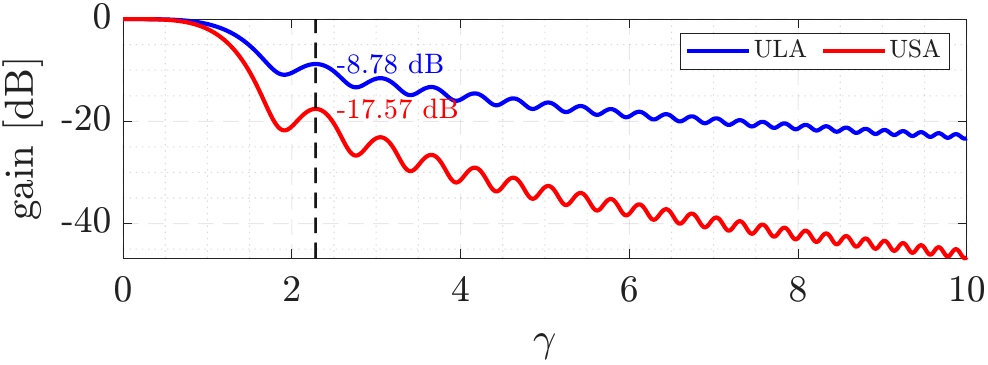}
\setlength{\abovecaptionskip}{-5pt}
\caption{Normalized array gain for \ac{ULA} and \ac{USA} with respect to $\gamma$.}
\label{fig:ULA_USA_array_gain}
\end{figure}

The \ac{PSL} of a uniformly weighted array is determined by the secondary maxima of the gain function. As illustrated in Fig. \ref{fig:ULA_USA_array_gain} (blue curve), the Fresnel-based gain function $\GULA(\gamma)$ in \eqref{eqn-B6-17} attains its global maximum at $\gamma = 0$, and is monotonically decreasing. The location of the first sidelobe is obtained by differentiating $\GULA(\gamma)$ with respect to $\gamma$, and solving for its roots. The first positive root occurs at $\gamma_{\mathrm{PSLL}} \approx 2.28$, where the gain in \eqref{eqn-B6-17} evaluates to $\GULA(\gamma_{\mathrm{PSLL}})\approx 0.1323$, yielding $\mathrm{PSLL} \approx 10\log_{10}(0.1323) \approx -8.7\,\mathrm{dB}$.

\begin{figure}[t]
\centering
\includegraphics[width=1\columnwidth]{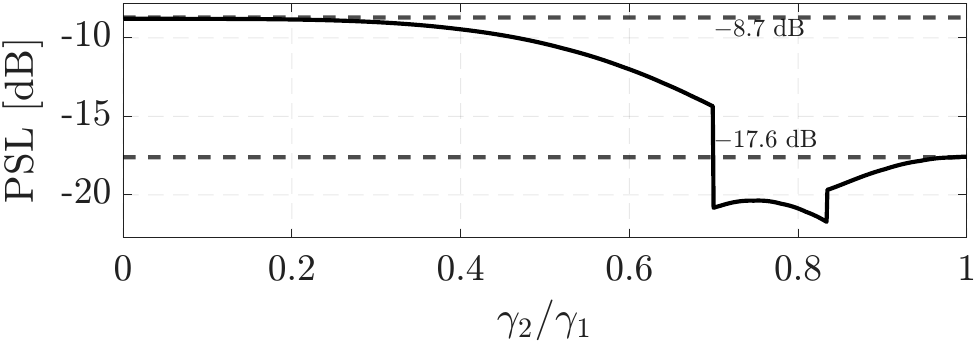}
\setlength{\belowcaptionskip}{-10pt}
\setlength{\abovecaptionskip}{-5pt}
\caption{Peak sidelobe levels with respect to $\hat{\eta}=\tfrac{\gamma_2}{\gamma_1}$.}
\label{fig:gamma_ratio_URA}
\end{figure}

\subsection{Uniform Rectangular Array}
We consider a \ac{URA} comprising $N = N_1 N_2$ antenna elements, where $N_1$ and $N_2$ denote the number of antenna elements along the $y$-axis and $z$-axis, respectively. The inter-element spacing along both axes is uniform and set to $d = \tfrac{\lambda}{2}$. The width-to-height ratio of the \ac{URA} is defined as $\eta = \tfrac{N_2}{N_1}$. By varying $\eta$, different array configurations can be realized. In particular, a \ac{USA} is obtained when $\eta = 1$, while a \ac{ULA} corresponds to $\eta \gg 1$ or $\eta \ll 1$. The aperture length of the \ac{URA}, derived from the Pythagorean theorem, is given by $D = d \sqrt{N_1^2 + N_2^2}$. For brevity, we omit the intermediate steps in deriving the array gain. The final expression for the array gain of a \ac{URA} is given in terms of Fresnel functions as \cite{ahmed2025NF}
\begin{equation}
\GURA(\gamma_1,\gamma_2) = \frac{\left[C^2(\gamma_1) + S^2(\gamma_1)\right]\left[C^2(\gamma_2) + S^2(\gamma_2)\right]}{(\gamma_1 \gamma_2)^2},
\label{eqn_IIIB_3}
\end{equation}
where $\gamma_{1} = \sqrt{\tfrac{N_{1}^{2} d^{2} \beta_1}{2\lambda} r_\mathrm{eff}}$ and $\gamma_{2} = \sqrt{\tfrac{N_{2}^{2} d^{2} \beta_2}{2\lambda} r_\mathrm{eff}}$, with $\beta_1 = 1-\sin^{2} \theta \sin^{2} \varphi$ and $\beta_2 = 1-\cos^{2} \theta$. The gain expression in \eqref{eqn_IIIB_3} is the product of two one-dimensional Fresnel functions, each having the same structure as the \ac{ULA} pattern in \eqref{eqn-B6-17}.

To emulate \ac{URA} configurations ranging from a \ac{ULA} to a \ac{USA}, we evaluate \eqref{eqn_IIIB_3} for $\hat{\eta} = \tfrac{\gamma_2}{\gamma_1} \in (0,1]$. The corresponding \ac{PSL} values are computed for each configuration, as shown in Fig.~\ref{fig:gamma_ratio_URA}. For analytical clarity, we focus on the boresight case to highlight the impact of array geometry. In this scenario, $\hat{\eta} = \sqrt{\tfrac{N_2^2 \beta_2}{N_1^2 \beta_1}}$, which simplifies at boresight $(\varphi=0, \theta=\tfrac{\pi}{2})$ to $\hat{\eta} = \tfrac{N_2}{N_1}$. As illustrated in Fig.~\ref{fig:gamma_ratio_URA}, when $\hat{\eta} < 0.3$, the \ac{PSL} converges to $\unit[-8.7]{dB}$, consistent with the \ac{ULA} case. Conversely, when $\hat{\eta} = 1$, corresponding to a \ac{USA}, the \ac{PSL} decreases to $\unit[-17.6]{dB}$. Furthermore, for $\hat{\eta} \in [0.7, 0.83]$, the \ac{PSL} falls below $\unit[-17.6]{dB}$; however, the edges of the mainlobe for these specific configurations exhibit minor distortions and warrant further investigation. It is noteworthy that the \textbf{beamdepth is minimized for a \ac{ULA} and maximized for a \ac{USA} \cite{ahmed2025NF,10443535}. In contrast, the \ac{PSL} exhibits the opposite trend, attaining its minimum for a \ac{USA} and maximum for a \ac{ULA}}.

\subsection{Uniform Circular Array}
\begin{figure}[t]
\centering
\includegraphics[width=.8\columnwidth]{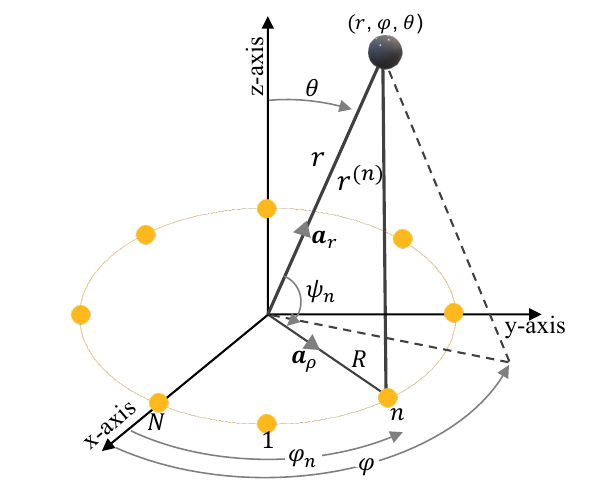}
\setlength{\belowcaptionskip}{-10pt}
\caption{A \ac{UCA} in the $xy$-plane and a \ac{NF} \ac{UE} is located $(r,\varphi, \theta)$.}
\label{uca_system_model}
\end{figure}

  We consider a \ac{UCA} of radius $R$ with $N$ antenna elements located in the $xy$-plane, as shown in Fig.~\ref{uca_system_model}. In polar coordinates, the $\nth{n}$ element is positioned at $(R,\varphi_n)$ with $\varphi_n = \tfrac{2\pi n}{N}$. The arc length between adjacent elements is $d = R \Delta \varphi = R \tfrac{2\pi}{N}$, which gives the UCA radius as $R = \tfrac{N d}{2\pi}$ \cite{hussain2025uniform}. The \ac{UE} is located at a distance $r$ from the center of the array, subtending azimuth angle $\varphi$ and elevation angle $\theta$. 
The distance between \ac{UE} and the $\nth{n}$ antenna element, as illustrated in Fig.~\ref{uca_system_model}, is given by
\begin{align}
    &r^{(n)} 
    = \sqrt{ r^2 + R^2 - 2 r R \cos(\psi_n) }, \label{eq:uca_cosine}\\ 
    &\overset{(a)}{=} \sqrt{ r^2 + R^2 - 2 r R \sin\theta \cos(\varphi - \varphi_n) }, \notag \\
    &\overset{(b)}{\approx} r - R \sin \theta \cos(\varphi - \varphi_n) + \tfrac{R^2}{2r} \Big( 1 - \sin^2\theta \cos^2(\varphi - \varphi_n) \Big).
    \label{eq:uca_general}
\end{align}
Here, (a) follows from $\cos(\psi_n) = \hat{a}_\rho \cdot \hat{a}_r = (\hat{a}_x \cos \varphi_n + \hat{a}_y \sin \varphi_n) \cdot (\hat{a}_x \sin\theta \cos\varphi + \hat{a}_y \sin\theta \sin\varphi + \hat{a}_z \cos\theta) = \sin\theta \cos(\varphi - \varphi_n)$ \cite{balanis2016antenna}. Furthermore, (b) is obtained by applying a second-order Taylor expansion around $r$, following the approach in~\eqref{eqn-A4}. The $\nth{n}$ component of the normalized \ac{NF} array response vector for a \ac{UCA}, based on \eqref{eq:uca_general}, is given by
{\begin{equation}\small
\mathbf{b}_n(\varphi,\theta,r) \approx \tfrac{1}{\sqrt{N}} 
e^{-j\nu \!\left( R \sin \theta \cos(\varphi - \varphi_n) 
- \tfrac{R^2}{2r} \big( 1 - \sin^2\theta \cos^2(\varphi - \varphi_n) \big)\right)}.
\label{eqn-NF_UCA}
\end{equation}
\normalsize}
The array gain for the \ac{UCA} is obtained by substituting \eqref{eqn-NF_UCA} into the general definition of the axial pattern defined in \eqref{eqn-A7r}
{\begin{equation}\small
\GUCA(\varphi,\theta,r) 
= \left| \frac{1}{N} \sum_{n=1}^N 
e^{\,j \tfrac{2\pi}{\lambda} \tfrac{R^2}{2}r_\mathrm{eff} 
\left( 1 - \sin^2\theta \cos^2(\varphi-\varphi_n) \right)} \right|^2,
\end{equation}
\normalsize}
where $r_\mathrm{eff}=\big|\tfrac{r-\rf}{r\,\rf}\big|$. For analytical tractability, we define $\zeta=\tfrac{\pi}{\lambda}\tfrac{R^2}{2}\,r_\mathrm{eff}\sin^2\theta$. Substituting $R=\tfrac{\DUCA}{2}$, where $\DUCA$ denotes the aperture diameter, and the Rayleigh distance $\RDUCA=\tfrac{2\DUCA^2}{\lambda}$, this simplifies to $\zeta=\tfrac{\pi\RDUCA}{16}\,r_\mathrm{eff}\sin^2\theta$. Furthermore, by exploiting the inherent rotational symmetry of the \ac{UCA}, we set $\varphi=0$ without loss of generality, yielding the simplified array gain 
\begin{equation}
\GUCA(\theta,r) \approx \left|\frac{1}{N} \sum_{n=1}^{N} e^{-2j \zeta \cos^2(\varphi_n)} \right|^2.
\end{equation}
Using the trigonometric identity $\cos(2x) = 2\cos^2(x)-1$, 
\begin{equation}
\GUCA(\theta,r) \approx \left|\frac{1}{N} \sum_{n=1}^{N} e^{-j \zeta \cos(2\varphi_n)} \right|^2.
\end{equation}
For large $N$, the summation approaches the integral \cite{balanis2016antenna}
\begin{equation}
\frac{1}{N} \sum_{n=1}^{N} e^{-j \zeta \cos(2\varphi_n)}=\frac{1}{2\pi}\int_0^{2\pi} e^{-j\zeta \cos(\varphi)} \, d\varphi = J_0(\zeta),
\end{equation}
where $J_0(\zeta)$ is the zeroth-order Bessel function of the first kind. Hence, the axial pattern of the \ac{UCA} is approximated as
\begin{equation}
\GUCA(\theta,r) \approx \left| J_0(\zeta) \right|^2, \ \text{where } \quad \zeta=\tfrac{\pi\RDUCA}{16}\,r_\mathrm{eff}\sin^2\theta.
\label{eqn_UCA_beam_pattern}
\end{equation}
To compute \ac{PSL} for the \ac{UCA}, we utilize the following asymptotic expansion of the Bessel function~\cite{olver1954asymptotic}:
\begin{equation}
J_0(\zeta) \sim \sqrt{\tfrac{2}{\pi \zeta}} \cos\!\big(\zeta - \tfrac{\pi}{4}\big).
\label{eq:J0_asymp}
\end{equation}
The maxima occur when the cosine argument is an integer multiple of $\pi$, i.e., $\zeta-\tfrac{\pi}{4}=k\pi \;\Longrightarrow\; \zeta = k\pi+\tfrac{\pi}{4},\; k \in [0,1,2,\ldots]$. The first sidelobe corresponds to $k=1$, which yields $\zeta_{\mathrm{PSLL}} = \tfrac{5\pi}{4}$. At this point, the gain function evaluates to $|J_0(\zeta_{\mathrm{PSLL}})|^2 = \tfrac{8}{5\pi^2} \approx 0.162$, resulting in a \ac{PSL} for the \ac{UCA} as $\mathrm{PSLL} \approx 10\log_{10}(0.162) \approx -7.9~\mathrm{dB}$.

\begin{figure}[t]
\centering
\includegraphics[width=.8\columnwidth]{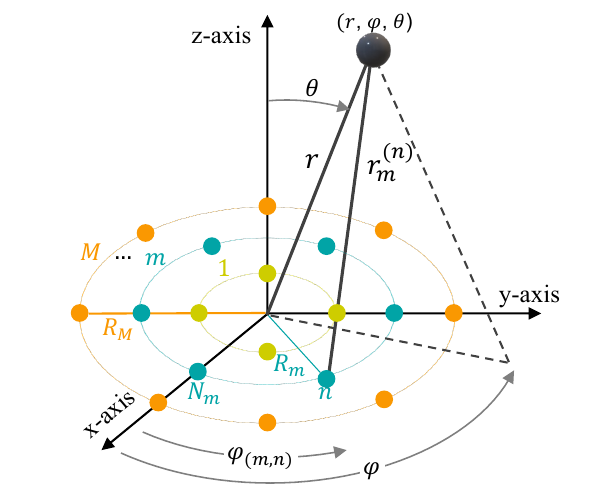}
\setlength{\belowcaptionskip}{-10pt}
\caption{A \ac{UCCA} in the $xy$-plane and a \ac{NF} \ac{UE} is located $(r,\varphi, \theta)$.}
\label{ucca_system_model}
\end{figure}

\subsection{Uniform Concentric Circular Array}
We consider a \ac{UCCA} comprising $M$ concentric rings, as shown in Fig. \ref{ucca_system_model}. The radius of the $\nth{m}$ ring is $R_m$ and contains $N_m$ isotropic elements uniformly spaced at angles $\varphi_{(m,n)}=2\pi n/N_m$. The total element count is $N=\sum_{m=1}^M N_m$. The radius of $\nth{m}$ \ac{UCCA} ring is assumed to be proportional to the index $m$, i.e., $R_{m}=\frac{m}{M} R_{M}$ and $N_{m}=\frac{m}{M} N_{M}$, where $R_{M}$ and $N_{M}$ denote the radius and number of elements of the outermost ring \cite{10517927}. The \ac{NF} array response vector for the $\nth{(m,n)}$ element of the \ac{UCCA} is given by \eqref{eqn-NF_UCA}, with the subscript $(n)$ replaced by $(m,n)$. The full array response vector is obtained by concatenating the array response vectors of all $M$ rings. Finally, the corresponding array gain for the \ac{UCCA} follows by substituting the \ac{NF} array response vector into the general definition of the axial pattern in \eqref{eqn-A7r}
\begingroup
\begin{equation}
\small
\begin{aligned}
\GUCCA(\varphi,\theta,r)
&= \left|\frac{1}{N} \sum_{m=1}^M \sum_{n=1}^{N_m} 
\mathrm{e}^{\,j \tfrac{2\pi}{\lambda}\tfrac{R_m^2}{2}\, r_{\mathrm{eff}}
\big(1-\sin^2\theta\cos^2(\varphi-\varphi_{(m,n)})\big)}\right|^2,  \\
&\stackrel{(a)}{=} \left|\frac{1}{N} \sum_{m=1}^{M} 
\mathrm{e}^{\,j \tfrac{2\pi}{\lambda}\tfrac{R_{m}^{2}}{2}\big(1-\tfrac{1}{2}\sin^{2}\theta\big) r_{\mathrm{eff}}}
\right. \\
&\quad\;\;\left.\times \sum_{n=1}^{N_{m}} 
\mathrm{e}^{\,-j \tfrac{2\pi}{\lambda}\tfrac{R_{m}^{2}\sin^{2}\theta}{4}\cos\big(2(\varphi-\varphi_{(m,n)})\big)\,r_{\mathrm{eff}}}
\right|^2,  \\
&\stackrel{(b)}{\approx} \left|\frac{1}{N}\sum_{m=1}^{M} 
\mathrm{e}^{\,j\,2\zeta_m\big(\tfrac{1}{\sin^{2}\theta}-\tfrac{1}{2}\big)}\,N_m J_0(\zeta_m)\right|^2.
\end{aligned}
\label{eq:UCCA_gain}
\end{equation}
\endgroup
In (a), we apply the identity $\cos^2 x = \tfrac{1}{2}(1+\cos 2x)$ to separate the terms. 
Then, in (b), we define $\zeta_m \triangleq \tfrac{\pi}{\lambda}\tfrac{R_m^2}{2}\,r_{\mathrm{eff}}\sin^2\theta$. 
When the ring radii are chosen as $R_m=\tfrac{m}{M}\tfrac{\DUCCA}{2}$, and with $\RDUCCA=\tfrac{2\DUCCA^2}{\lambda}$ denoting the Rayleigh distance of the outermost ring, this reduces to $\zeta_m = \tfrac{\pi \RDUCCA}{16} \tfrac{m^2}{M^2}\, r_{\mathrm{eff}} \sin^2\theta$. Furthermore, $J_0(\zeta_m)$ is obtained from the summation over the $\nth{m}$ ring, and its derivation is similar to the \ac{UCA} case, as explained in the previous subsection. 

The gain expression in \eqref{eq:UCCA_gain} for the \ac{UCCA} is intractable due to the summation over $m$. To simplify, we consider the boresight case, $\theta=0^\circ$, where $J_0(0)=1$, and replace the summation with an integral, yielding the following expressions:
\begin{align}
\GUCCA(\varphi,\theta,r)\Big|_{\theta=0}
{\approx} &\left|\frac{N_M}{M N}\int_{0}^{M} m\,e^{-j\frac{\xi}{M^2}m^2}\,dm\right|^2, \notag \\
\overset{(a)}{\approx}& \left|\frac{N_M}{M N}\cdot\frac{M^2}{2j\xi}\big(1-e^{-j\xi}\big)\right|^2, \notag \\
\overset{(b)}{\approx}& \left|\frac{\sin(\xi/2)}{\xi/2}\right|^2,
\label{eq:UCCA_gain-1}
\end{align}
where $\xi = \tfrac{\pi \RDUCCA}{8} r_{\mathrm{eff}}$. In (a) we apply $\int m\,e^{j a m^2}\,dm = \frac{1}{2 j a} e^{j a m^2}$. We set $\tfrac{N_M M}{N}=1$ in (b) and $\lvert 1-e^{-j\xi}\rvert^2 = 4\sin^2(\xi/2)$ simplifies to the sinc form.

\begin{figure}[t]
\centering
\includegraphics[width=0.8\columnwidth]{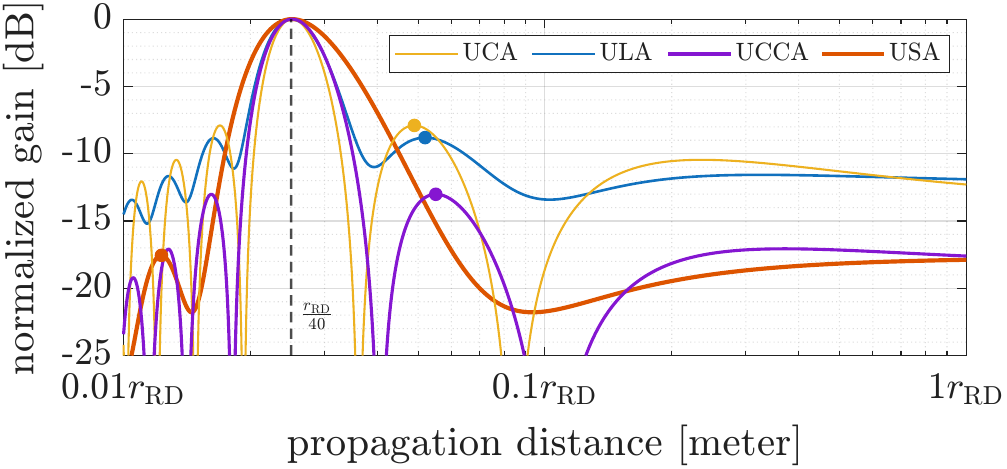}
\caption{Axial beam patterns for \ac{ULA}, \ac{USA}, \ac{UCA}, and \ac{UCCA}, with the \ac{NF} beam focused at $\tfrac{\RD}{40}$.}
\label{fig_PSLL_range}
\end{figure}

The \ac{PSL} for the \ac{UCCA} is determined from the secondary maxima of the gain function in \eqref{eq:UCCA_gain-1}, given by $\GUCCA(\xi) \approx \left|\tfrac{\sin(\xi/2)}{\xi/2}\right|^2$. The mainlobe attains its peak value of unity at $\xi=0$, while the first sidelobe occurs at $\xi \approx 3\pi$. Evaluating the gain at this point gives $\GUCCA(\xi_{\mathrm{PSLL}}) \approx \left(\tfrac{1}{3\pi/2}\right)^2 \approx 0.045$, which yields $\mathrm{PSLL} \approx 10\log_{10}(0.045) \approx -13.46~\mathrm{dB}$.

\begin{figure}[t]
\centering
\includegraphics[width=0.8\columnwidth]{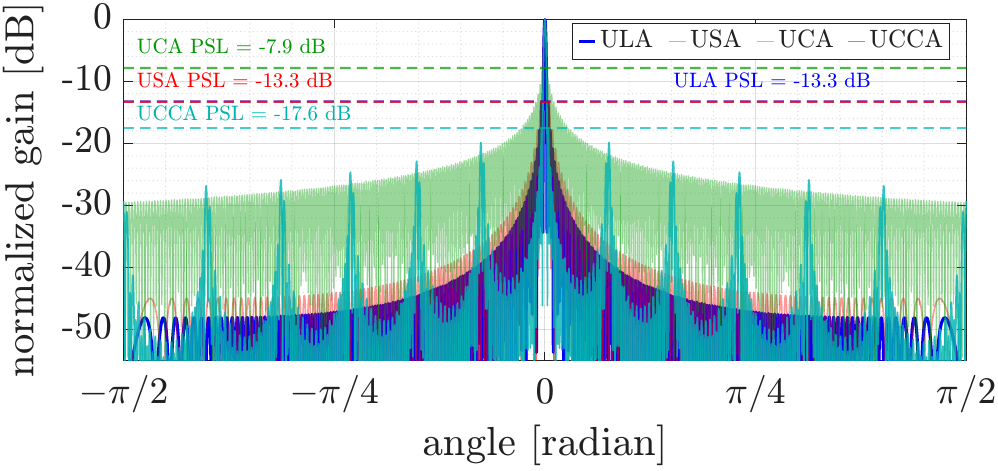}
\setlength{\belowcaptionskip}{-20pt}
\caption{Angular beam patterns for \ac{ULA}, \ac{USA}, \ac{UCA}, and \ac{UCCA}, with the \ac{NF} beam focused at azimuth angle $\varphi=0$.}
\label{fig_PSLL_ang}
\end{figure}

\section{Simulation Results}\label{Sec-IV}
In this section, we compare the \ac{SLL} in terms of \ac{PSL} and \ac{ISL}, and validate the theoretical analysis through numerical simulations. Subsequently, we compare the achievable sumrate across the array geometries.
\subsection{Sidelobe level vs. Array Geometry}
We consider \ac{ULA}, \ac{USA}, \ac{UCA}, and \ac{UCCA}, each with an aperture length of $D=\unit[1.26]{m}$ and operating at $f=\unit[15]{GHz}$. Since the aperture length is fixed, the Rayleigh distance is identical across all configurations and equals $\RD \approx \unit[162]{m}$. Industry trends indicate that aperture length is a more critical constraint than the number of antenna elements. For example, Qualcomm recently unveiled a 4096-element Giga-\ac{MIMO} prototype operating at $\unit[13]{GHz}$ within a form factor comparable to existing 256-element \ac{5G} \acp{BS} \cite{qualcomm2024mwc}, illustrating the broader trend of integrating large-scale arrays into practical form factors.

The axial patterns for the array configurations are shown in Fig.~\ref{fig_PSLL_range}, with the \ac{NF} beam focused at a distance of $\tfrac{\RD}{40}$ and at boresight for all geometries except the \ac{UCA}. For the \ac{UCA}, the coplanar direction is considered, as it does not support \ac{NF} beams in the boresight direction. The selected focusing directions also correspond to the best-case \ac{PSL} for each array configuration.

Unlike the \ac{FF} case, where the mainlobe is symmetric around the steered angle, the \ac{NF} mainlobe is asymmetric around the focal distance $\rf$. The left and right $3$-dB points of the mainlobe are denoted by $\rf^{\min}$ and $\rf^{\max}$, respectively. In practice, the right interval, $\rf^{\max}-\rf$, is larger than the left interval, $\rf-\rf^{\min}$. However, when expressed in terms of reciprocal distances, these intervals become symmetric, i.e., $\abs{\rf^{\max}-\rf} > \abs{\rf-\rf^{\min}}$ but $\lvert\tfrac{1}{\rf^{\max}} - \tfrac{1}{\rf}\rvert = \lvert\tfrac{1}{\rf} - \tfrac{1}{\rf^{\min}}\rvert$. This asymmetry is not evident in Fig.~\ref{fig_PSLL_range} due to the logarithmic scaling of the range axis. The sidelobes preceding the mainlobe are termed \emph{forelobes}, while those following it are termed \emph{aftlobes}. Forelobes and aftlobes differ in both width and number, with forelobes generally being more numerous and of higher magnitude. In general, forelobes primarily interfere with \ac{NF} \acp{UE}, whereas aftlobes affect \ac{FF} \acp{UE}.

\begin{table}[t]
\centering
\renewcommand{\arraystretch}{1.1}
\caption{\ac{PSL} and \ac{ISL} in range and angle domains.}
\begin{tabular}{
 >{\centering\arraybackslash}m{1.5cm} | 
 >{\centering\arraybackslash}m{1.2cm} >{\centering\arraybackslash}m{1.2cm} |
 >{\centering\arraybackslash}m{1.2cm} >{\centering\arraybackslash}m{1.2cm}
}
\toprule
\multirow{3}{*}{\textbf{Geometry}} & 
\multicolumn{2}{c}{\textbf{PSLL [dB]}} & 
\multicolumn{2}{c}{\textbf{ISLL [dB]}} \\
\cmidrule(lr){2-3} \cmidrule(lr){4-5}
 & \textbf{Range} & \textbf{Angle} & \textbf{Range} & \textbf{Angle} \\
\midrule
UCA & -7.9 & -7.9 & -0.4 & 1.9 \\
\hline
ULA & -8.7 & -13.3 & -1.3 & -9.6 \\
\hline
UCCA & -13.4 & \textbf{-17.6} & -7.2 & \textbf{-10.4} \\
\hline
USA & \textbf{-17.6} & -13.3 & \textbf{-12.1} & -9.6 \\
\bottomrule
\end{tabular}
\label{tab:psl_isl_comparison}
\end{table}

{In the range domain, the \ac{USA} achieves the lowest \ac{PSL} of $\unit[-17.6]{dB}$, followed by the \ac{UCCA}, \ac{ULA}, and finally the \ac{UCA}}. However, as shown in Fig.~\ref{fig_PSLL_ang}, in the angle domain, the \ac{UCCA} attains the lowest \ac{PSL} of $\unit[-17.6]{dB}$, followed by the \ac{USA} and \ac{ULA}, each with $\unit[-13.3]{dB}$. The \ac{UCA} yields the worst performance across both range and angle domains, with a \ac{PSL} of $\unit[-7.9]{dB}$. Table~\ref{tab:psl_isl_comparison} summarizes the \ac{PSL} and \ac{ISL} values computed numerically. Furthermore, the \ac{USA} achieves the lowest \ac{ISL} in the range domain, while the \ac{UCCA} exhibits superior performance in the angle domain. It is noteworthy that the \ac{PSL} remains invariant with respect to the number of antenna elements $N$; increasing $N$ does not reduce the \ac{PSL}. In contrast, the \ac{ISL} for each array configuration can be further reduced by increasing $N$.

\subsection{Achievable Sumrate}
 We perform Monte Carlo simulations to compare the achievable sumrate across different array configurations under the constraint of a fixed aperture length. The carrier frequency is set to $f = \unit[15]{GHz}$, with an aperture length of $D=\unit[1.26]{m}$ and a Rayleigh distance of $\RD \approx \unit[162]{m}$. The number of antenna elements $N$ for each array geometry is selected to achieve an equal aperture length. Specifically, for the \ac{ULA}, $N=128$; for the \ac{UCA}, $N=400$; for the \ac{UCCA}, the outermost ring comprises $N_M=400$ elements with $M=40$ rings, yielding a total of $N=\sum_{m=1}^M N_m = 8200$ elements; and for the \ac{USA}, $N=7225$. For each array configuration, five \acp{UE} are uniformly distributed within the range interval $ \mathcal{U}(2D, \RD)$.

\begin{figure}[t]
\centering
\includegraphics[width=1\columnwidth]{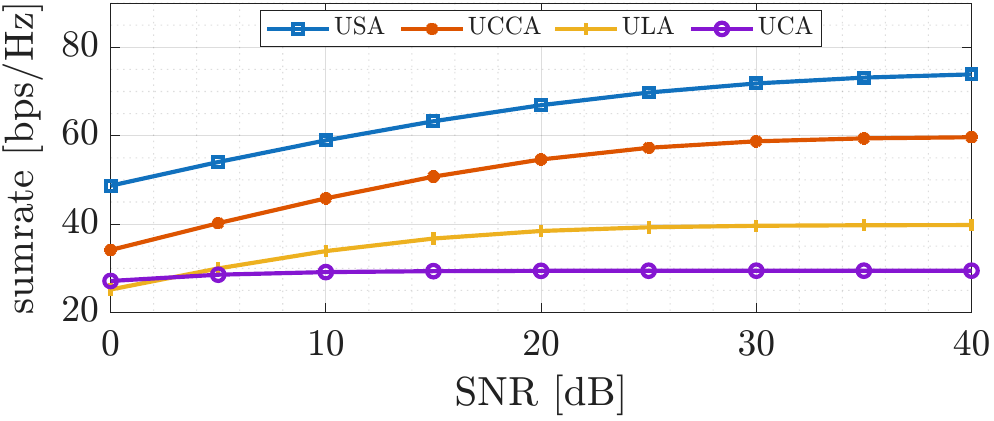}
\setlength{\belowcaptionskip}{-10pt}
\setlength{\abovecaptionskip}{-11pt}
\caption{Achievable sumrate vs. \ac{SNR} for \ac{ULA}, \ac{USA}, \ac{UCA}, and \ac{UCCA}.}
\label{fig_sumrate_sameD}
\end{figure}

Fig.~\ref{fig_sumrate_sameD} shows the achievable sumrate with respect to \ac{SNR} for all array geometries. Unsurprisingly, the \ac{USA} achieves the highest sumrate, followed by the \ac{UCCA}, \ac{ULA}, and finally the \ac{UCA}. The superior performance of the \ac{USA} stems from its lower \acp{SLL}, which ensures that different \acp{UE} experience minimal inter-user interference. Furthermore, the sumrate results are consistent with the \ac{SLL} comparison across different geometries in the previous subsection, thereby validating the impact of \acp{SLL} on sumrate performance.

\section{Conclusion} \label{Sec-V}
In this paper, we analyzed how \acp{ASL} vary with array geometry. We derived array gain expressions for uniform planar arrays and used them to characterize the \ac{PSL}. Our analysis demonstrated that the \ac{USA} yields the lowest \acp{ASL}, followed by the \ac{UCCA}, \ac{ULA}, and \ac{UCA}. Numerical simulations further confirmed that the \ac{USA} attains the highest sumrate, consistent with the \acp{SLL} analysis. As future work, windowing methods may be explored for suppressing \acp{ASL} \cite{ahmed_slepian}. Furthermore, flexible antenna arrays may be investigated to reduce \acp{SLL} across both axial and lateral dimensions simultaneously.

\bibliographystyle{IEEEtran}
\bibliography{Bibliography/IEEEabrv,Bibliography/my2bib}
\end{document}